\documentclass[aps,pra,reprint,superscriptaddress,floatfix,showpacs]{revtex4-1}
\usepackage{xcolor}
\usepackage{cancel}

\usepackage{float}
\usepackage{empheq} 
\usepackage{braket}
\usepackage{multirow}

\usepackage{bm}

\usepackage{graphicx}
\usepackage{hyperref}
\usepackage{svg}
\usepackage{natbib}
\usepackage{siunitx}
\usepackage[normalem]{ulem}

\hypersetup{
    colorlinks,
    linkcolor={blue!50!blue},
    citecolor={blue!50!blue},
    urlcolor={blue!
    80!black}
}

\begin{document}
\title[A universal and stable metasurface for photonic quasi bound state in continuum coupled with two dimensional semiconductors]{A universal and stable metasurface for photonic quasi bound state in continuum coupled with two dimensional semiconductors} 

\author{Brijesh Kumar}
\author{Anuj Kumar Singh}
\author{Kishor K Mandal}
\author{Parul Sharma}
\author{Nihar Ranjan Sahoo}
\author{Anshuman Kumar}
\email{anshuman.kumar@iitb.ac.in}
\affiliation{Laboratory of Optics of Quantum Materials, Department of Physics, IIT Bombay, Mumbai - 400076, India}
\date{\today}

\begin{abstract}
Strong coupling of excitons to optical cavity modes is of immense importance to understanding the fundamental physics of quantum electrodynamics at the nanoscale as well as for practical applications in quantum information technologies. There have been several attempts at achieving strong coupling between excitons in two dimensional semiconductors such as transition metal dichalcogenides (TMDCs) and photonic quasi- bound states in the continuum (BICs). We identify two gaps in the platforms for achieving strong coupling between TMDC excitons and photonic quasi-BICs: firstly, in the studies so far, different cavity architectures have been employed for coupling to different TMDCs. This would mean that typically, the fabrication process flow for the cavities will need to be modified as one moves from one TMDC to the other, which can limit the technological progress in the field. Secondly, there has been no discussion of the impact of fabrication imperfections in the studies on strong coupling of these subsystems so far. In this work, we address these two questions by optimizing a cavity with the same architecture which can couple to the four typical TMDCs (MoS$_2$, WS$_2$, MoSe$_2$, WSe$_2$) and perform a detailed investigation on the fabrication tolerance of the associated photonic quasi-BICs and their impact on strong coupling.
\end{abstract}

\maketitle

\emph{Introduction--} In recent years, photonic bound states in the continuum (BIC) have attracted significant attention both for their fundamental physics as well as important applications\cite{Hsu2013,Azzam2020,PhysRevLett.100.183902}. A BIC in a lossless system has an infinite quality factor, with no radiative loss to the environment. BICs are broadly classified into two types, namely, symmetry protected BICs and parameter tuning based Friedrich-Wintgen BICs\cite{Hsu2016}. In practice, however, in order to excite the BIC modes, it becomes necessary to open the radiation channel slightly, thereby making the Q-factors to be finite. A common way of achieving this is deviating slightly from the perfect BIC condition, via a small modification in the geometry of the system and it can be shown that $Q\propto\alpha^{-2}$, where $\alpha$ is the geometrical asymmetry factor\cite{Koshelev2018}. These so called quasi-BICs present a versatile platform for enhanced light matter interaction and have been shown to be relevant for numerous applications in the development of nanoscale light sources\cite{Huang2020,Kodigala2017}, sensors\cite{Hsiao2022,Meudt2020,Romano2019} and nonlinear optical devices\cite{PhysRevLett.121.033903,PhysRevB.97.224309,Hwang2022}.

On the other hand, two dimensional semiconductors such as monolayer transition metal dichalcogenides (TMDCs), owing to the high binding energy, oscillator strength and gate tunability of excitons have emerged as a promising platform for various optoelectronic applications at room temperature\cite{Mak2016,Xia2014}. Strong coupling of excitons to photonic modes is of great relevance to understanding the fundamental physics of quantum electrodynamics at the nanoscale as well as for practical applications in quantum information technologies\cite{Schneider2018,Liu2014,AlAni2022}. Typically, one requires the rate of coherent interaction between the two subsystems to exceed the dissipation rate to observe such a strong coupling\cite{Liu2014}. Several routes have been explored for achieving strong coupling between excitons in TMDCs and photonic quasi-BICs\cite{Cao2020,PhysRevB.98.161113,PhysRevB.105.165424,Kravtsov2020,AlAni2021,Qin2021,Wu2022}. However, in these efforts so far, there are two gaps in the platforms for achieving strong coupling between TMDC excitons and photonic quasi-BICs. First of all, different nanophotonic cavity architectures have been employed for coupling to the excitons of different TMDCs, which have different exciton resonances and linewidths. For practical implementation, this would require the fabrication process flow for the cavities to be modified as one moves from one TMDC to the other, which can limit the technological progress in the field.  Secondly, the impact of fabrication imperfections on strong coupling of these subsystems has not been studied so far.
This is an extremely important practical consideration since for example, despite the promise of photonic quasi-BICs, in practice the achieved Q-factors are typically smaller than those observed in traditional whispering gallery based resonators or photonic crystal cavities\cite{Khne2021,Meng2022}. In this work, we address these two gaps by optimizing a universal metasurface design with the same architecture which can couple to the four typical TMDCs (MoS$_2$, WS$_2$, MoSe$_2$, WSe$_2$). For the first time, using statistical methods, we perform a detailed investigation on the fabrication tolerance and stability of the associated photonic quasi-BICs and their impact on strong coupling.

\begin{figure*}[htbp]
  \centering
  \includegraphics[width = 2\columnwidth]{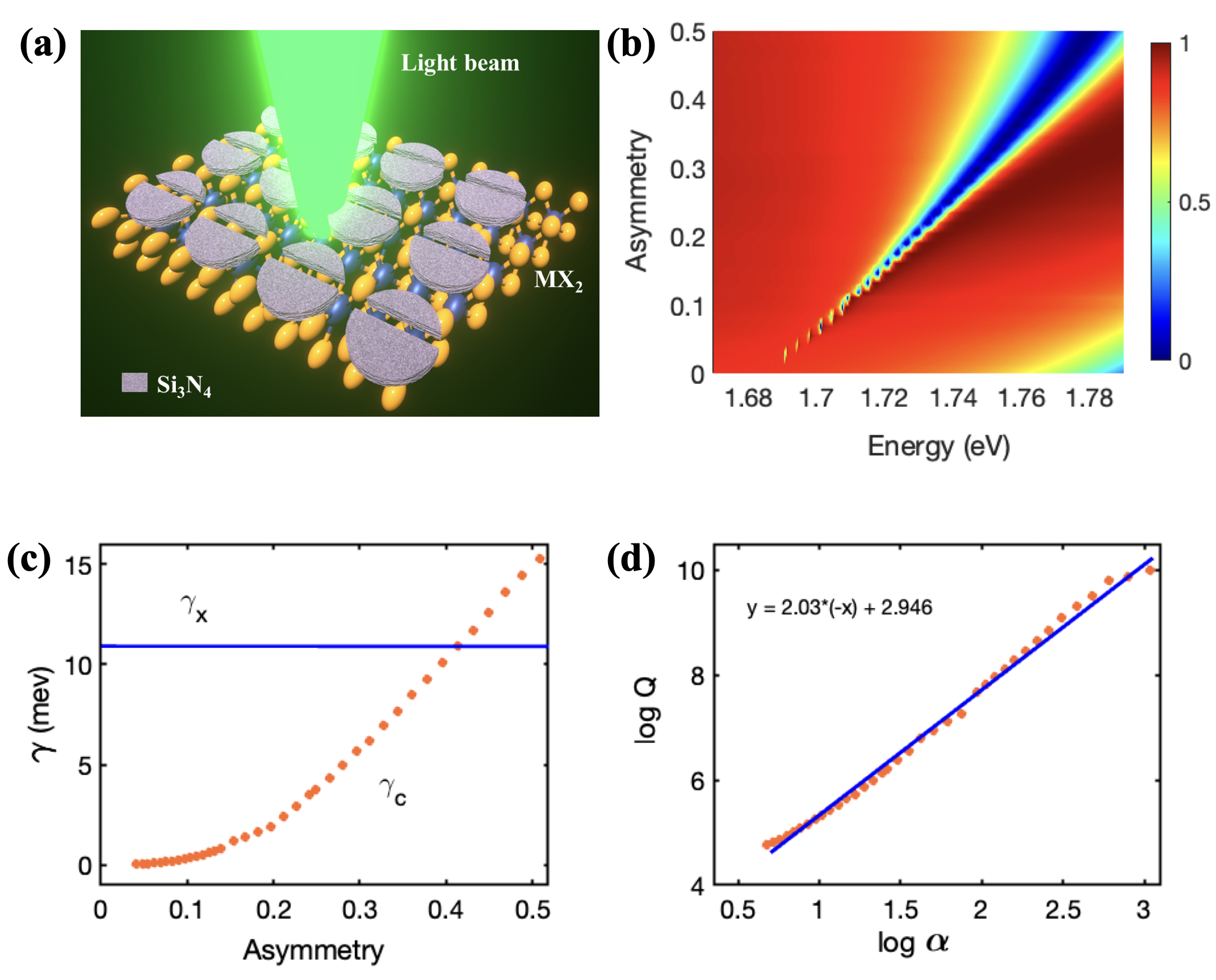}
  \caption{\textbf{Bound State in continuum in floating split disk structure}- (a) The schematic of the quasi-BIC cavity-emitter system includes a monolayer of MX$_2$ as an exciton emitter and a metasurface of split disk cavity supporting quasi-BIC. (b) The transmission spectra are plotted as a function of the asymmetry parameter $\alpha= 1- \frac{\sigma_1}{\sigma_0}$  (c) Cavity radiation loss \textbf{$\gamma_c$} as a function of asymmetry of disk defined as $\alpha$.  (d) Cavity Q-factor as a function of asymmetry of disk}
  \label{fig1}
\end{figure*}
\emph{Analysis of cavity modes---} We consider a cavity architecture as shown in Fig.~\ref{fig1}(a). As discussed above, we convert true BICs into quasi-BICs with a finite Q-factor by breaking the in-plane symmetry of a unit cell in y-direction. Initially, we slotted the disk of radius ($r$) with a rectangular bar (of dimension $2r \times [y_{min}+y_{\text{max}}]$) with $y_{min}=y_{\text{max}} = 20 nm$. This rectangular bar divides the disk into equal parts. Subsequently, $y_{\text{max}}$ is varied to above 20 nm to tune the asymmetry parameter. We define the asymmetry parameter as $\alpha = 1 - \frac{\sigma_1}{\sigma_0}$, where $\sigma_0$ and $\sigma_1$ are the areas of the lower and upper part of the slotted disk. In order to characterize the quasi-BIC mode, we calculate the transmission spectrum for normally incident $x$-polarised light using finite element method (FEM) based full-wave electrodynamic simulations via \textsc{comsol multiphysics}\cite{multiphysics1998introduction} as shown in Fig.~\ref{fig1}(b). We use a low-loss CMOS compatible material, namely, silicon nitride (Si$_3$N$_4$) for the cavity. In our wavelength range of interest, its refractive index is around 2.02\cite{Yulaev2022,Luke:15}.

A quasi-BIC is manifested in the reflection and transmission spectra as a Fano resonance\cite{PhysRevLett.121.193903}. The transmission $T$ of a periodic photonic structure of isotropic dielectric permittivity can be related to Fano lineshape using explicit expansion of the Green's function of a non-Hermitian system\cite{RevModPhys.82.2257,PhysRevB.90.245140}
\begin{equation*}
  T(E)=T_{b}-T_0 \frac{\left(1+\frac{E-E_0}{\gamma q}\right)^2}{1+\left(\frac{E-E_0}{\gamma}\right)^2}
  \label{eq1}
\end{equation*}
    where $E_0$ and $\gamma_c$ are the resonance energy and linewidth respectively, $T_b$ represents the background contribution of nonresonant modes, $T_0$ is the transmission offset, $q$ is the Fano parameter, which becomes ill-defined for a true BIC as the mode becomes completely dark\cite{PhysRevLett.121.193903}. The transmission spectra are plotted as a function of the asymmetry parameter $\alpha$ in Fig.~\ref{fig1}(b), from which one can clearly see that a BIC is found at the symmetric structure ($\alpha = 0$) and transforms {\color{black}to} quasi-BIC with increasing asymmetry. Using Fano fitting as described above, the linewidth of this quasi-BIC is calculated and shown to increase with asymmetry ($\alpha$), as plotted in Fig.~\ref{fig1}(c). We further confirm the universal quadratic relation between Q-factor and asymmetry for the quasi-BIC ($Q\propto\alpha^{-2}$) in Fig.~\ref{fig1}(d). This equation is the quasi-BIC benchmark and valid for all-dielectric and plasmonic implementations\cite{Koshelev2018}.

\emph{Multipolar decomposition of the quasi-BIC mode---} Multipole decomposition method (MDM), which decomposes a given field distribution into a superposition of the fields produced by a set of multipoles, is a powerful technique to analyze BICs.  The inverse radiation lifetime $\gamma_r$ of such a system can be computed as\cite{Koshelev2019} as: 
\begin{equation}
    \frac{\gamma_r}{c} = |D_x|^2+|D_y|^2 \\
    \label{eq2}
\end{equation}
where $D_x$ and $D_y$ are the coupling amplitudes for the two orthogonal polarizations, which can be expressed in terms of multipoles as\cite{Evlyukhin2016}:
\begin{flalign}
     D_{x,y} = -\frac{\sqrt{2}\pi}{\lambda a}\biggl[p_{x,y} + \frac{ik_0}{c}T_{x,y}- \frac{1}{c}m_{y,x} &\\ -\frac{ik_0}{c}M_{yz,xz} -\frac{ik_0}{6}Q_{xz,yz} + ...\biggr] &
    \label{Eq3}
\end{flalign}
where $p_x$, $T_x$ are $x$ components of electric dipole and toroidal dipole moment respectively, $m_y$ is the $y$ component of the magnetic dipole moment; $Q_{xz}$ and $M_{yz}$ are the next higher order of multipoles, namely, electric and magnetic quadrupole moments. For the ideal BIC case (no asymmetry), it can be shown that $D_i$'s identically vanish resulting in zero radiative decay rate\cite{Koshelev2019}. For the $\Gamma$- BIC case, the {\color{black}multipoles} have no contribution to the far field along the $z$ direction. At other {\color{black}$k-$}points, the multipoles are either in phase or in antiphase depending on the $k$-vector and structural parameters. If the sum of all these multipoles goes to zero, this leads to an accidental BIC.
\begin{figure}[htbp]
  \centering
  \includegraphics[width = \columnwidth]{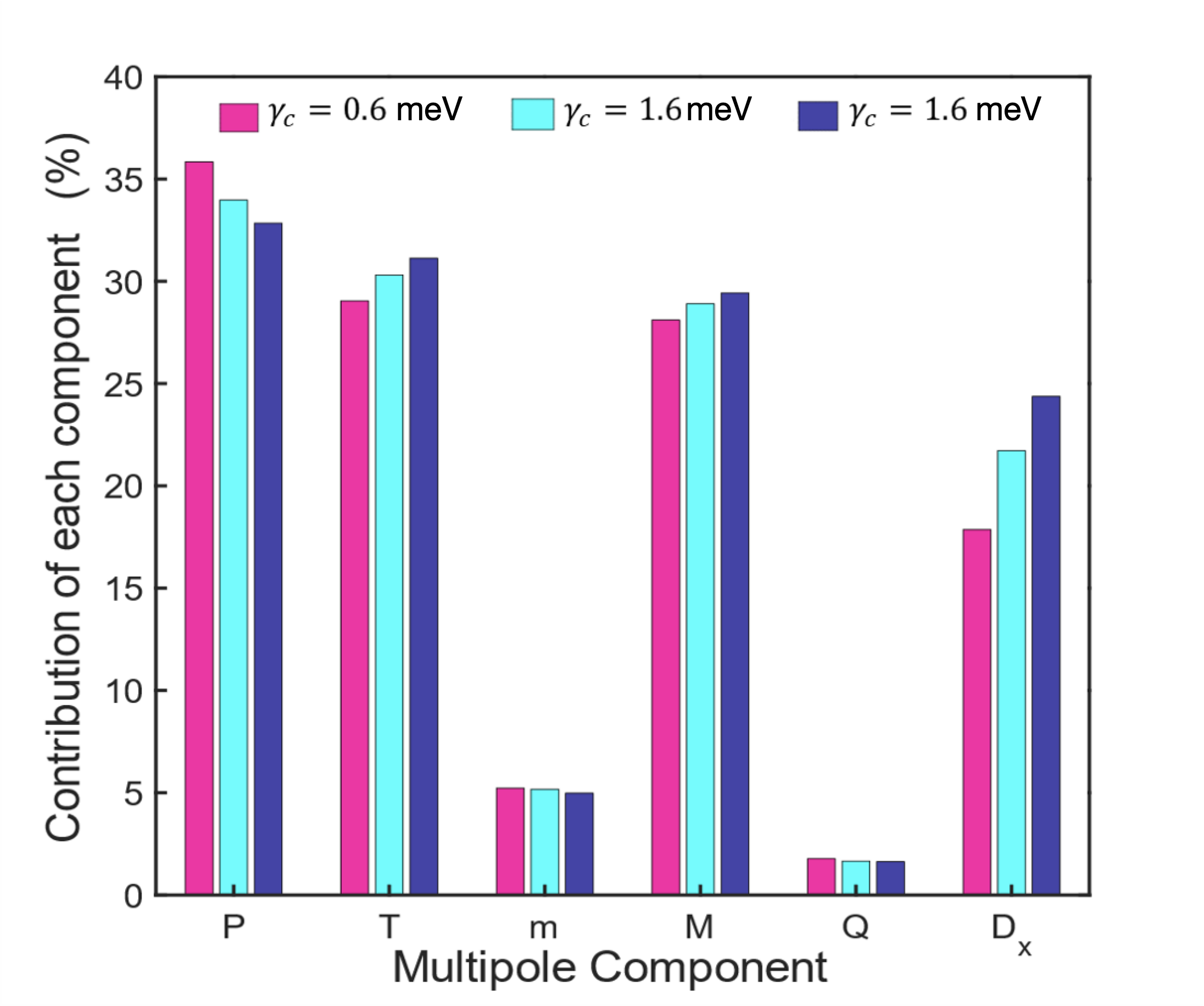}
  \caption{\textbf{Multipole analysis}-Normalised multipolar decomposition amplitude contribution in quasi-BIC at different cavity losses: $\gamma_c =0.6$ (black), $\gamma_c = 1.6 $, (red) $\gamma_c =3.5 $ meV (blue) respectively. We include five multipolar components, electric dipole moments \textbf{P}, magnetic dipole moments \textbf{m}, toroidal dipole moment \textbf{T}, electric and magnetic quadrupole moment \textbf{Q,M}  and $D_x$ as given in Eq. \ref{Eq3} respectively. \textcolor{black}{$D_x$ is increasing with the cavity loss, consistent with Eq. \ref{eq2}}} 
  \label{fig2}
\end{figure}

\begin{figure*}[htbp]
  \centering
  \includegraphics[width = 2\columnwidth]{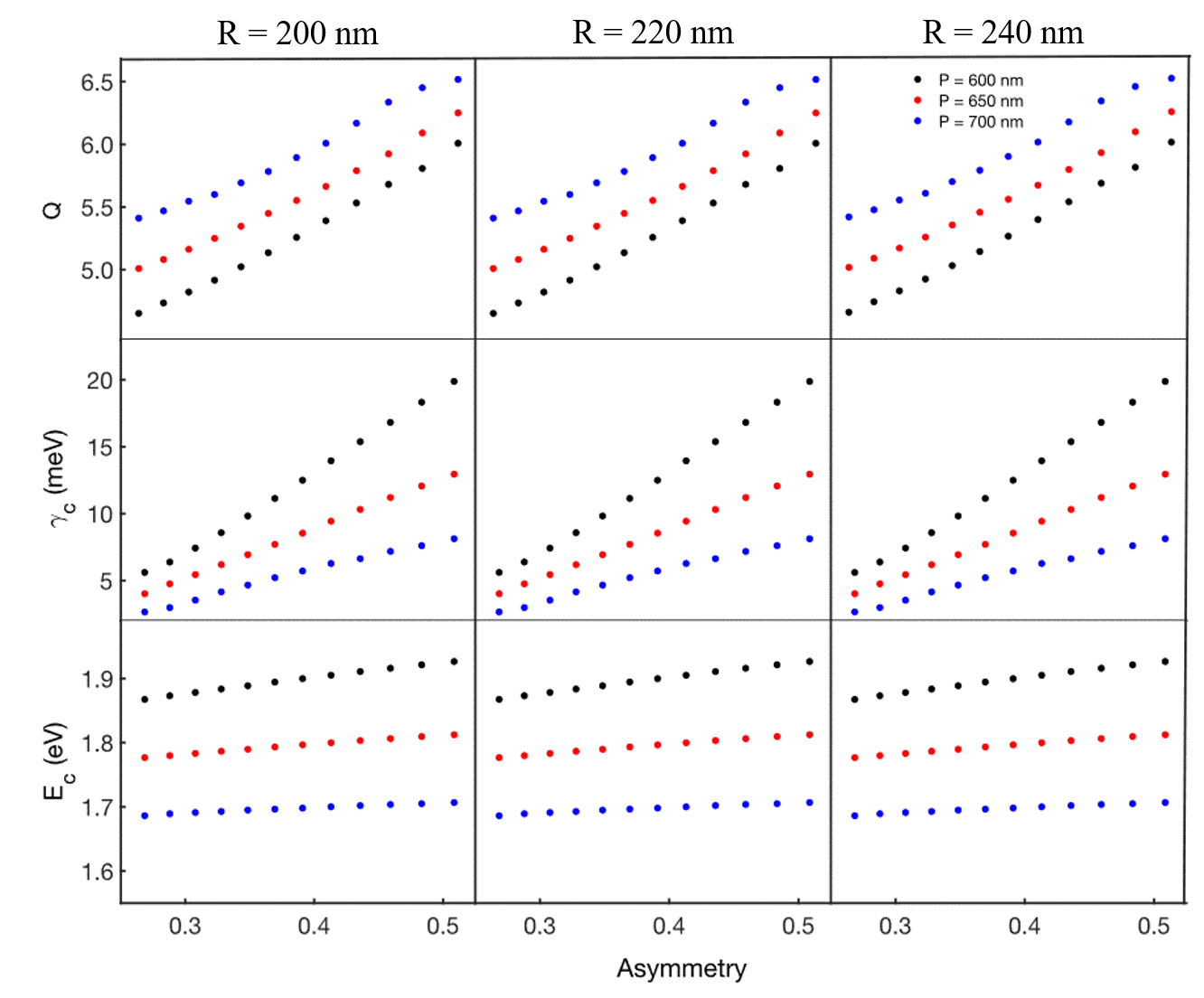}
  \caption{\textbf{Stability analysis of quasi-BIC mode}- Q-factor, cavity radiation, loss, and cavity excitation peak is measured as a function of asymmetry for nine combinations of periodicity [\textbf{600} (black) \textbf{650} (red) \textbf{700} (blue) \textbf{nm}] and radius [\textbf{200}  \textbf{220}  \textbf{240} \textbf{nm}]. quasi-BIC condition $Q \propto \alpha^{-2}$ is consistent for all combination}
  \label{fig3}
\end{figure*}

We calculate all the multipole moments using the definition given in \cite{Evlyukhin2016} by volume integration over the unit cell. These multipoles are plotted in Fig.~\ref{fig2}. In the top row of Fig.~\ref{fig2}, we present the multipole contributions for three different radiation loss rates of the cavity in increasing order, that is, $\gamma_c = 0.6, \ 1.2, \ 3.5$ meV. From the figure, it is clear that electric dipole, toroidal and magnetic quadrupoles are the dominant multipoles contributing to transmission and reflection peaks. \textcolor{black}{{In the supporting information, we show how the broadening of these peaks results in a corresponding broadening of the reflection spectra and the respective electric field profiles at the dip in transmission indicating an enhancement in light confinement as the quasi-BIC approaches the BIC state.}} 
\newline
\emph{Stability of quasi-BIC---}
The multipole decomposition theory in the above section explains the existence of a quasi-BIC mode in our structure. In this section, we test the robustness of this quasi-BIC as a function of structural parameters. We choose different radii and periodicity of the structure to see the stability of BIC-mode to mimic fabrication imperfections. Fig.~\ref{fig3}(a-c) shows that the inverse quadratic relation remains valid for all radii at constant periodicity. The average deviation in slope is $0.01 /nm$ for radius change. For periodicity change, the slope changes at a rate of $10^{-3}/nm$
From this test, it can be concluded that this quasi-BIC can be observed in practical devices where fabrication imperfections are inevitable.
\newline
\emph{Strong coupling in TMDCs using quasi-BICs---}
Having prepared the high Q-factor quasi-BIC based on the dielectric metasurface, a TMDC monolayer is positioned on the top of the Si$_3$N$_4$ slotted disk to produce a strong coupling between the exciton and quasi-BIC cavity mode at room temperature. We first employ a WSe$_2$ monolayer to illustrate the strong coupling enabled by quasi-BIC. The thickness of WSe$_2$ is taken as 0.7 nm\cite{AlAni2021}. The anti-crossing characteristic in the absorption spectrum reveals strong light-matter interaction, resulting in cavity polaritons. We further consider other TMDC monolayers on the top of the slotted disk cavity. These cavity designs, while of the same topology, are re-optimized to show quasi-BICs at the exciton frequencies of the different TMDCs. The complex dielectric permittivity of monolayer TMDCs as a function of the photon energy $E$ is given by\cite{Miroshnichenko2010}:
\begin{equation*}
    \epsilon(E)=\epsilon_{\mathrm{B}}+\frac{f}{E_{0}^2-E^2-i \Gamma E}
\end{equation*}
Where $\epsilon_B$ is the background permittivity, $E_0$, $f$ and $\Gamma$ denote the resonance energy, oscillator strength, and the damping rate of oscillator. These TMDC parameters are taken from\cite{AlAni2021}. As shown in Fig.~\ref{fig1}(a), the cavities are optimized to guarantee a perfect overlap of the cavity spectrum with the corresponding TMDCs monolayer exciton. To do this, we note that the monolayer exciton resonance energy is fixed, so we can vary the cavity resonance $E_Q$ by sweeping the height of the designed quasi-BIC cavity resonator. 
\begin{figure*}[htbp]
  \centering
  \includegraphics[width = 2\columnwidth]{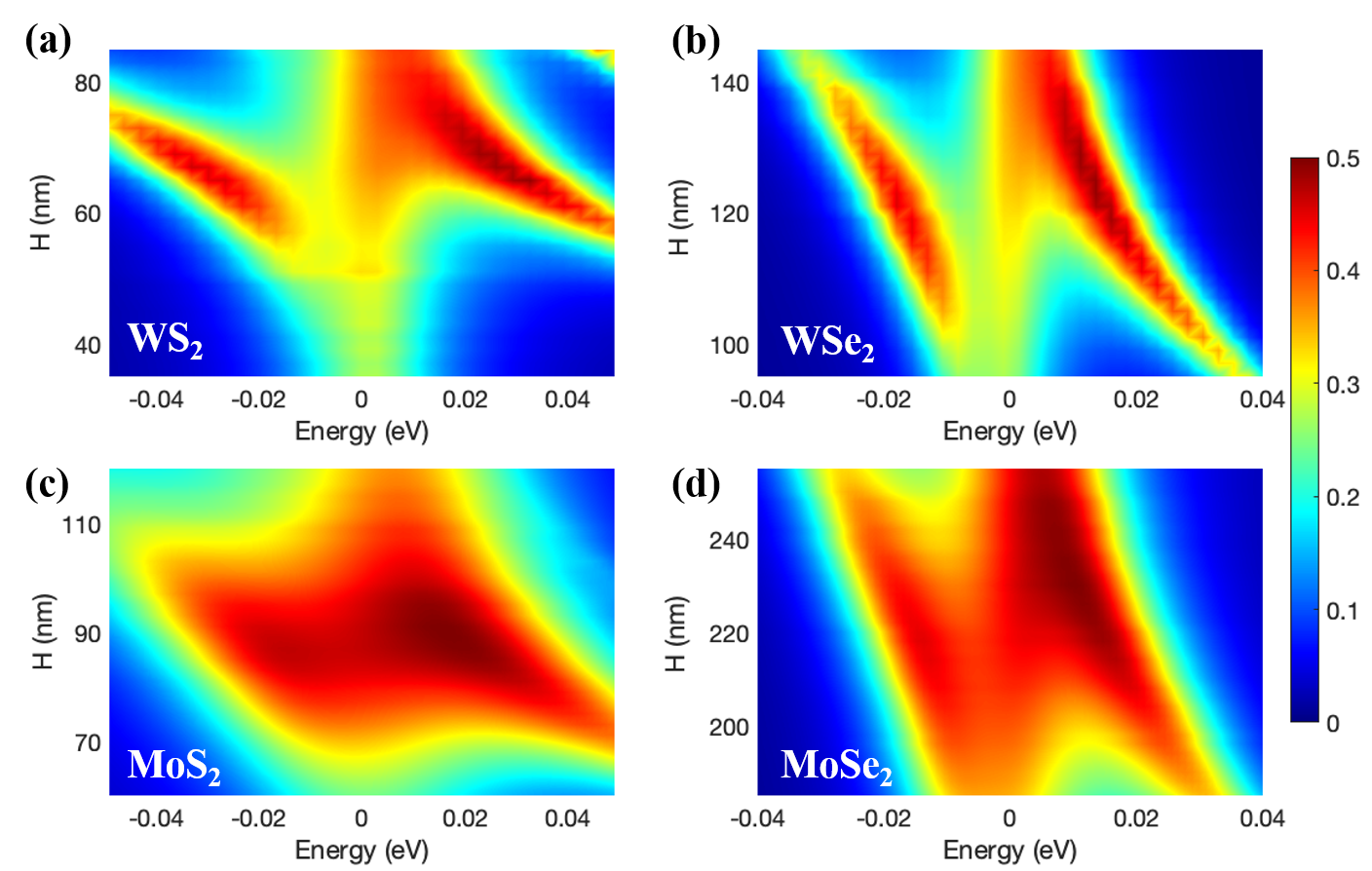}
  \caption{\textbf{Strong critical coupling between TMDC exciton and Quasi-BIC}. (a-d) Absorption spectrum for all TMDCs {WSe$_2$, WS$_2$,MoSe$_2$, MoS$_2$} for perfectly matched damping rates at the avoided crossing. Detuning is obtained via varying the height of the cavity.}
  \label{fig4}
\end{figure*}

\textcolor{black}{In our calculations we swept the height of the cavity while keeping the asymmetry parameter fixed, to sweep the detuning. We find that height variation only weakly affects the cavity damping rate, which is mostly determined by the asymmetry parameter. Such a tunability of the linewidth and the resonance frequency offers unprecedented ability to study different regimes of the coupling between excitons and optical modes, ranging from strong coupling, critical coupling and weak coupling, exploring the entire phase diagram of coupled light matter system\cite{Zanotto2014}.}
Since our quasi-BIC slotted cavity is composed of a dielectric material with a negligible imaginary part of refractive index, the damping rate of cavity arises from the radiative loss alone. On the other hand, the absorption of hybrid system -- quasi-BIC cavity integrated with the TMDC monolayer -- results from the dissipation rate of the TMDC only. To show the power of this platform, we present the calculated absorbance for the case of matched damping rate between cavity and monolayer in Fig(\ref{fig4}e-g) for four TMDCs. In this setup, the so called strong critical coupling can be achieved. We show that this functionality can be easily achieved with the same topology of the cavity for all four typical TMDC materials, which not only have different resonant frequencies but also different damping rates. To perform this optimization, the damping rate of the cavities was extracted from the Fano fitting in accordance with Eq.~\ref{eq1}.  In the color map, energy is shown on the x-axis, the cavity height on y-axis and absorption on colour bar. The anti-crossing nature behaviour seen in the colour map is signature of strong coupling. The fact that the absorbances reach the value of 0.5 shows that we are also in the critical coupling regime for all four materials\cite{Zanotto2014}. It should be noted that critical coupling is important because in this case all of the incoming energy can be perfectly absorbed by the system. The optimised geometrical parameters of the cavity for realizing strong critical coupling with the four typical TMDCs are shown in Table~\ref{tabel:1}.

\begin{table}
\caption{Optimized parameters of the cavity for polaritonic coupling with different TMDC excitons. 
\label{tabel:1}}
\begin{tabular}{|c|c|c|c|c|}
\hline
\multicolumn{1}{|l|}{Cavity parameter} &\multicolumn{1}{l|}{WS$_2$} & \multicolumn{1}{l|}{{MoS$_2$}} & \multicolumn{1}{l|}{WSe$_2$} & \multicolumn{1}{l|}{MoSe$_2$}\\ \hline
          Periodicity (nm)  & 550 & 600   & 665       & 665                         \\ 
              Radius  (nm) &  250  &   250 &  200       &  200                     \\ 
               $y_{\text{max}}$ (nm) & 150 &182 & 120 & 118                          \\ 
               Height (nm) & 65 & 87 & 112 & 219\\ \hline
\end{tabular}
\end{table}

\begin{figure*}[htbp]
  \centering
  \includegraphics[width = 2\columnwidth]{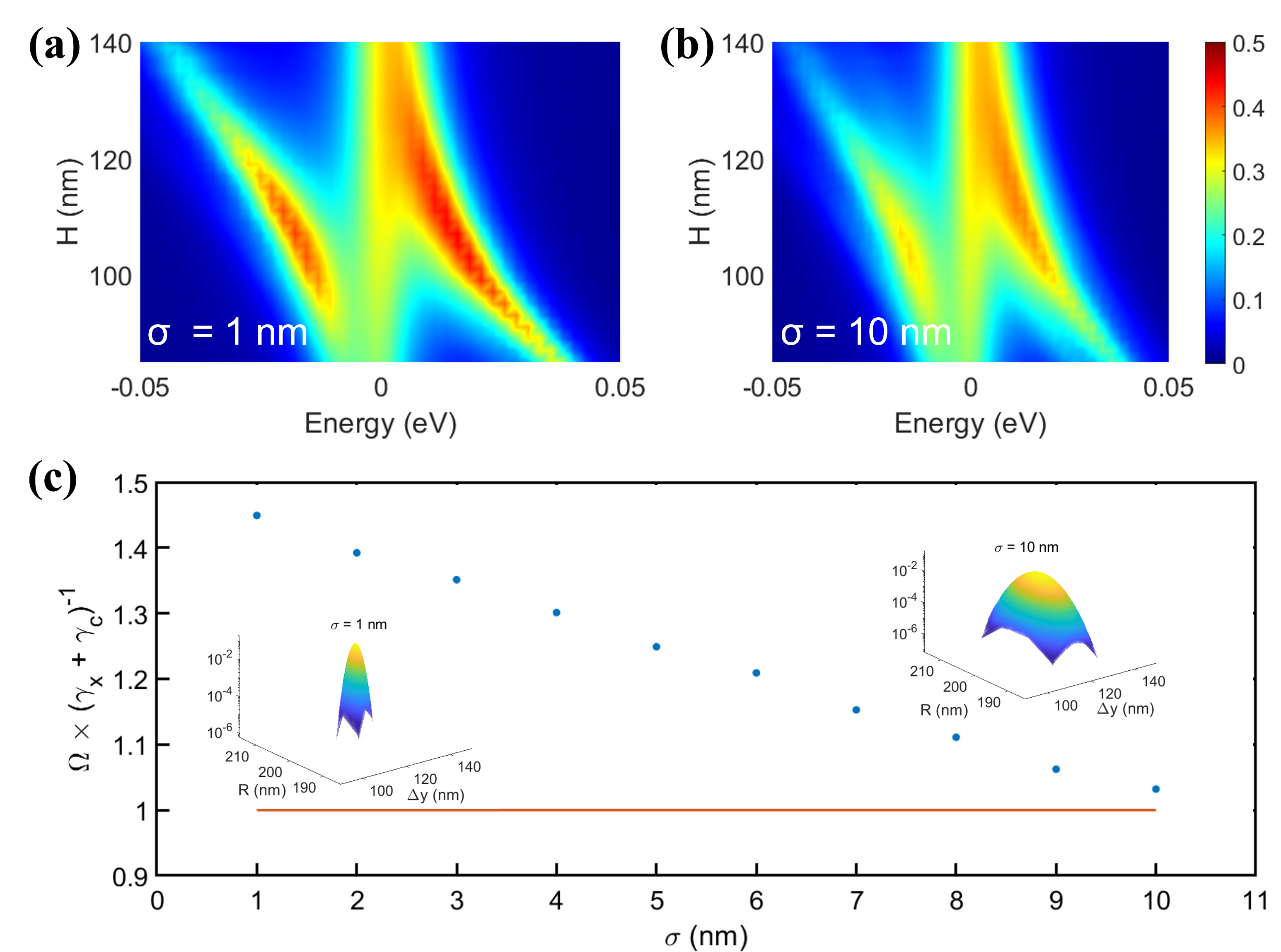}
  \caption{\textbf{Fabrication robustness in strong coupling:} (a -- b) \textcolor{black}{strong polariton splitting at different variations. (c) Rabi splitting condition ($\frac{\Omega}{\gamma_x + \gamma_c}\ge 1$) as a function of variation of cavity geometry ($\sigma$).} }
  \label{fig5}
\end{figure*}

\textcolor{black}{\emph{Influence of fabrication imperfections on strong coupling ---} Typical lithographic techniques for creating the nanostructures for realizing the quasi-BIC modes suffer from inevitable fabrication imperfections. It is therefore important to consider the impact of fabrication imperfections on the strong coupling behaviour of our exciton-cavity system. We choose two geometric parameters of the cavity, which have a significant influence on the quasi-BIC cavity resonance frequency and linewidth. Electron beam lithography (EBL) and subsequent liftoff process induced variations in radius and asymmetry parameter are chosen for this study. A brute force way to model the effect of imperfect fabrication on optical response is to introduce geometrical variation across several unit cells of the metasurface. However, this method needs a large simulation region over hundreds or thousands of unit cells. Here instead, we simulate the strong coupling using a single unit cell at different geometrical parameters and calculate the optical response using statistical methods. The two parameters, namely, the radius and $y_{\text{max}}$ are varied $7.5\%$ and $18\%$ respectively. \textcolor{black}{These numbers correspond to a radius variation of $\pm 15 $ nm and a $y_{\text{max}}$ variation of $\pm 20$ nm}. We construct a five-by-five square grid with a centre at the optimized parameter. Absorption spectra for these 25 points were simulated by \textsc{comsol} and linearly interpolated in \textsc{matlab} to form a fine grid of $1000 \times 1000$. Then, $N = 5000$ random points are sampled from a Gaussian distribution to mimic the fabrication-induced variation shown in Fig.~\ref{fig5}. Spectra corresponding to all random points is then averaged to calculate the absorption of the imperfectly fabricated metasurfaces. In Fig.~\ref{fig5}, we show a colour plot with energy detuning and height on $x \, \ y$-axes and absorption on the colour axis. These spectra are calculated for different variations of the Gaussian distribution, namely, $\sigma$. The plots show that strong coupling is stable up to a variation of 10 nm, but the absorption amplitude is reduced from 0.5 to 0.3. This shows that despite these fabrication imperfections, the system can stay in the strong coupling regime within this range of imperfections.}

\emph{Conclusion--} In summary, we have presented a cavity architecture which works well for strong coupling with all four typical TMDC excitons. The quasi-BIC optical mode itself is shown to be stable towards variations of geometrical parameters of the cavity, which is confirmed via explicit multipole moment calculations, where we show that the radiative damping rate of the cavity can be precisely engineered to match the exciton linewidth for critical coupling. We further show the robustness of this strong coupling regime towards inevitable fabrication imperfections of varying degree. The methodology demonstrated in our work should find wide applications in understanding not only the role of geometrical features of the cavity but also the practically important impact of fabrication imperfections on the strong coupling between quasi-BIC cavity modes and excitons in low-dimensional materials.

\section*{Acknowledgments}
B.K. acknowledges support from Prime minister's research fellowship (PMRF), Government of India. A.K. acknowledges funding support from the Department of Science
and Technology via the grants: SB/S2/RJN-110/2017,
ECR/2018/001485 and DST/NM/NS-2018/49.

\bibliographystyle{unsrt}  
\bibliography{references}

\end{document}